\documentclass[12pt,preprint]{aastex}

\slugcomment{Article for The Astronomical Journal}

\shorttitle{Interpretation of the Tadpole galaxy VV29}
\shortauthors{Djorgovski et al.}

\begin{document}

%% LaTeX will automatically break titles if they run longer than
%% one line. However, you may use \\ to force a line break if
%% you desire.

\title{Interpretation of the Tadpole VV29 Merging Galaxy System using
Hydro-Gravitational Theory}

%% Use \author, \affil, and the \and command to format
%% author and affiliation information.
%% Note that \email has replaced the old \authoremail command
%% from AASTeX v4.0. You can use \email to mark an email address
%% anywhere in the paper, not just in the front matter.
%% As in the title, you can use \\ to force line breaks.

\author{Carl H. Gibson\altaffilmark{1}}
\affil{Departments of Mechanical and Aerospace Engineering and
Scripps Institution
of Oceanography, University of
California,
      San Diego, CA 92093-0411}

\email{cgibson@ucsd.edu}

\and

\author{Rudolph E. Schild}
\affil{Center for Astrophysics,
      60 Garden Street, Cambridge, MA 02138}
\email{rschild@cfa.harvard.edu}

%% Notice that each of these authors has alternate affiliations, which
%% are identified by the \altaffilmark after each name.  Specify alternate
%% affiliation information with \altaffiltext, with one command per each
%% affiliation.

\altaffiltext{1}{Center for Astrophysics and Space Sciences, UCSD}

%% Mark off your abstract in the ``abstract'' environment. In the manuscript
%% style, abstract will output a Received/Accepted line after the
%% title and affiliation information. No date will appear since the author
%% does not have this information. The dates will be filled in by the
%% editorial office after submission.

\begin{abstract}
Hubble Space Telescope (HST/ACS) images of the galaxy merger Tadpole (VV29=Arp
188=UGC 10214) are interpreted using the hydro-gravitational
theory of Gibson 1996-2000 (HGT)
that predicts galaxy masses within about 100 kpc ($3.1 \times
10^{21}$ m) are dominated by dark halos of planetary mass
primordial-fog-particles (PFPs) in dark proto-globular-star-clusters (PGCs).
According to our interpretation, stars and young-globular-clusters (YGCs)
appear out of the dark as merging  galaxy components VV29cdef move
through the baryonic-dark-matter halo of the larger galaxy VV29a creating
luminous star-wakes.  Frozen PFP planets are evaporated by radiation and tidal
forces of the intruders. Friction from the gas accelerates an accretional
cascade of PFPs to  form larger planets, stars and YGCs of the filamentary
galaxy VV29b.    Star-wakes show that galaxy VV29c, identified as
a blue dwarf by radio telescope
observations of gas density and velocity  (Briggs et al. 2001), with companions
VV29def entered the dark halo of the larger VV29a galaxy at a radius $4
\times 10^{21}$ m and then spiraled in on different tracks toward frictional
capture by the VV29a core.  A previously dark dwarf
galaxy is identified from a Keck spectrographic study showing a VV29c
star-wake dense cluster of YGCs aligned to $1^o$ in a
close straight row
\citep{tra03}.
\end{abstract}

%% Keywords should appear after the \end{abstract} command. The uncommented
%% example has been keyed in ApJ style. See the instructions to authors
%% for the journal to which you are submitting your paper to determine
%% what keyword punctuation is appropriate.


%% From the front matter, we move on to the body of the paper.
%% In the first two sections, notice the use of the natbib \citep
%% and \citet commands to identify citations.  The citations are
%% tied to the reference list via symbolic KEYs. The KEY corresponds
%% to the KEY in the \bibitem in the reference list below. We have
%% chosen the first three characters of the first author's name plus
%% the last two numeral of the year of publication as our KEY for
%% each reference.

\section{Introduction}

The peculiar filamentary galaxy Tadpole (VV29=Arp 188=UGC 10214) at distance
$4 \times 10^{24}$ m and the similar merging galaxy system with a long luminous
tail Mice (NGC 4676) at distance $3 \times 10^{24}$ m were chosen to be among
the first images obtained by the HST/ACS Hubble Space Telescope, taking
advantage of the wide angle field of view of the newly installed Advanced
Camera for Surveys (April 30, 2002, press release) to provide better
information about the origins of these distant mysterious systems.  It was
speculated
\citep{tre01} that the long Tadpole tail might reflect the presence of a
nearby, non-baryonic, cold-dark-matter (CDM) ``Dark-Matter'' halo by forming a
tidal bridge of stars from VV29a extending toward a massive, but invisible,
CDM halo object.  However, ``CDM Halos'' are excluded by HGT cosmology
because non-baryonic dark matter (NBDM) by its weakly collisional nature is
highly diffusive with long mean free paths for collision so it cannot
possibly form the gravitationally bound  ``CDM Halo'' objects required
by CDM cosmologies.   According to HGT cosmology the NBDM diffuses to form
outer halos of large galaxies or galaxy clusters.  NBDM halos have smaller
densities than inner baryonic-dark-matter (BDM) galaxy halos even though they
have larger masses.  Sand et al. 2002 have constrained the density profile of
a lensing galaxy cluster within 100 kpc to a form that excludes the steep
universal CDM density profiles predicted by CDM cosmological simulations at
the 98$\%$ CL, strongly contradicting the CDM paradigm for galaxy formation.

The radio telescope measurements
\citep{bri01} clearly demonstrate the existence of a blue dwarf galaxy VV29c
embedded in the VV29a image with velocity different than the rapid VV29a
rotation velocity of about 330 km s$^{-1}$, so the massive CDM halo
interpretation of VV29b
\citep{tre01} becomes unnecessary. Briggs et al. 2001  propose a
different \citep{tt72} tidal tail interpretation of Tadpole where the blue
dwarf galaxy VV29c enters from the Northwest, experiences a close encounter
with VV29a, ejects (``slings'') the filamentary galaxy VV29b to the Southeast,
and then departs away from the observer along the VV29a line of sight to a
distance comparable to the length of VV29b about
$4 \times 10^{21}$ m.  The CDM-halo tidal bridge interpretation
\citep{tre01} is not excluded by the \citep{bri01} title, ``Did VV 29 collide
with a dark Dark-Matter halo?'' Both of these ``tidal tail'' interpretations
are based on the Toomre \& Toomre 1972   collisionless gravity tidal force
models, and  are inconsistent with HGT and with numerous previously unseen
details revealed by HST/ACS images such as Figure 1.

Thin tidal tails in frictionless tidal models are the result of thin
galaxy disks ejected so that by coincidence some portion is edgewise to the
observer, resulting in tidal lion or beaver tails that are bushy or
paddle-like at one or both ends \citep{mi01} rather than the thin and pointed
tadpole and mice tails observed in Tadpole, Mice and Antennae galaxy mergers
which we interpret as star wakes of merging galaxy cores moving through each
other's baryonic-dark-matter halos.  Random sprays of collisionless
particles in N body computer simulations must be  ignored in fitting tidal tail
simulations to galaxy merger observations.  No collisionless galaxy merger
simulation that matches the Tadpole geometry is presented or  referenced by
Briggs et al. 2001.  Furthermore, we see from the highest resolution HST/ACS
images that the disk of VV29c is perpendicular to the VV29b tail rather than
parallel as required, with VV29c embedded in VV29a, not far in the background
as proposed
\citep{bri01} assuming a frictionless galaxy interaction.  The VV29c galaxy
disk is intact and has not been  flung forward edgewise to
form VV29b as expected for a collisionless tidal tail model involving a
merging galaxy with a thin disk.

 From the HST/ACS images VV29c and its companions VV29def have been captured
by frictional forces of VV29a, and the star-wake VV29b cannot be described by
any version of the collisionless
\citep{tt72} ejected tidal tail model.  The oval features previously
interpreted as spiral arms of VV29a confute this pattern in the HST/ACS images
and appear instead as cylindrical luminous star-wakes.  The outermost
star-wake shows a spiral path that wraps once around VV29a and terminates with
VV29cf at half its initial distance from the VV29a center.  A smaller and
shorter star-wake splits into two  fragments VV29de about half way around
before these objects plunge toward VV29a at two thirds the way around and half
way to the  center.  One of the fragments VV29e apparently possesses an active
nucleus that  creates a spray of star formation extending $4
\times 10^{20}$ m to the Northeast, well away from the VV29a nucleus that lacks
any such AGN signature.  A similar spray of star formation from AGN radiation
into a  baryonic-dark-matter halo appears in a background galaxy
$2
\times 10^{21}$ m  on the plane due East of VV29a to illustrate this
phenomenon that is natural to assume from HGT but would otherwise be
mysterious.

The present frictional accretion merger and star wake interpretation of the
HST/ACS  Tadpole images is
explained by viscous-turbulent-diffusional hydro-gravitational fluid mechanics
that excludes CDM cosmologies and collisionless fluid mechanics assumptions
that are  the basis of
the frictionless flung disk \citep{bri01} and CDM dark halo
\citep{tre01} tidal tail interpretations.  In the following, we
briefly review the hydro-gravitational theory and the evidence
in its support and then compare its predictions with the HST/ACS
images.  Finally, some conclusions are provided.

\section{Hydro-Gravitational Theory Prediction of the Baryonic Dark Matter
Form}

Detailed evidence and analysis \citep{gs03} supporting the
hydro-gravitational theory (HGT) of self-gravitational structure formation
\citep{gib96}, and its prediction that the interstellar medium and
baryonic dark matter masses of galaxies should be dominated by
primordial-gas-planets (PFPs), are beyond the scope of the present paper. This
prediction of HGT was reached as a conclusion independently
\citep{sch96} from observed twinkling frequencies of galaxy lensed quasar
images, leading to the interpretation that the mass of the lensing galaxy was
dominated by ``rogue planets ... likely to be the missing mass''.
Length scales and acronyms relevant to self-gravitational structure formation
are given in Tables 1 and 2.  The Jeans length $L_J$ is the scale for which
pressure gradients in a self-gravitating ideal gas are smoothed by acoustic
wave propagation.  It is not a minimum scale for structure formation.  The
``pressure support'' argument used to bolster the Jeans 1902 gravitational
instability criterion fails because pressure forces depend on pressure
gradients, and these vanish by acoustic propagation on scales smaller than
$L_J$.  Non-acoustic density perturbations on scales smaller than the Jeans
length are absolutely unstable to gravitational structure formation unless
resisted by viscous forces, turbulence forces or diffusion of the fluid
particles.  Sub-Jeans scale self-gravitational instabilities appear in
numerical simulations in a stagnant gas
\citep{tru97}. However, these are dismissed as  numerical artifacts
(``artificial fragmentation'') based on the erroneous Jeans 1902 criterion,
and a Jeans scale digital filter is invented. Jeans
number digital filters (with Jeans number $J
\equiv
\Delta x / L_J \le 0.25$, where
$\Delta x$ is the cell size) are recommended versus numerical viscosity
normally used to suppress $L \le L_J$ instabilities in gravitohydrodynamic
simulations. We believe that the calculations prove the HGT claim that a
nearly stagnant gas is absolutely unstable to the formation of self-gravity
driven condensations and voids on predictable sub-Jeans scales (Table
1).

As indicated in Table 1, magnetic and other forces were negligible in the
viscous primordial plasma and buoyancy dominated gas \citep{gib00}.  The first
structures appeared at $L_{SV} \approx L_{ST}$ scales by gravitational
fragmentation in the plasma epoch, soon after the energy density of the hot
plasma matched the matter density ($\rho_{E} \approx
\rho_{M}
\approx 10^{-15}$ kg m$^{-3}$) at
$t \approx 7.5 \times 10^{11}$ s (25,000 years) with small
amplitude ($\delta
\rho /
\rho
\le 10^{-5}$) limited by the diffusive separation of the non-baryonic dark
matter to fill the voids and cut off the gravitational forcing.  The photon
viscosity was so large ($4 \times 10^{26}$ m$^2$ s$^{-1}$)
that the plasma turbulence was weak or non-existent.  The first
structures were at the horizon scale $L_H
\equiv ct
\approx L_{SV}
\approx L_{ST} \approx 3 \times 10^{20}$ m, where $c$ is the speed of light.
The baryonic density was $ \approx 3 {\--} 2 \times 10^{-17}$ kg m$^{-3}$ so
the first objects to form were protosupercluster fragments.    Further
fragmentation continued to protogalaxy masses, just before the plasma to gas
transition at 300,000 years. Turbulence was damped by buoyancy forces
at the Ozmidov scale $L_R \equiv [\varepsilon/N^3]^{1/2}$ of the
self-gravitational structures, with stratification frequency $N$,
starting at
$t \approx 10^{12}$ s (30,000 years) and the density and rate-of-strain at
that time were preserved as fossils of the earlier hydrodynamic state
\citep{gib99}, which determined the density of globular star clusters ($
\approx 10^{-17}$ kg m$^{-3}$),  $L_{SV}$, and the PFP masses.  CDM
condensations were impossible because $L_{SD_{CDM}} \gg L_H$.

Because the transition viscosity $\nu$ decreased by a factor of $\approx
10^{12}$, the fragmentation length scale decreased by a factor of $\approx
10^{6}$ (Table 1), giving a primordial gas fragmentation mass of $10^{-6}
M_{\sun}$ to produce primordial fog particles.  The gaseous
proto-galaxies also fragmented at the Jeans scale to form
proto-globular-cluster PGCs, but not for
the reasons suggested by Jeans 1902.
$L_J$ was considerably smaller than the proto-galaxy scale at decoupling, so
pressures could not equilibrate as gravitational structures formed.  The
initially isothermal gas developed  $L_J$ scale temperature fluctuations
smoothed by radiative heat transfer, giving
$L_J$ scale non-acoustic fragmentation sites.  Thus the proto-galaxies
fragmented into Jeans-mass PGC clouds of
primordial-fog-particle (PFP) fragments that comprise the baryonic-dark-matter
and ISM mass of all galaxies.  Most of the BDM ($\approx 97\%$)
remains dark, in $10^6 M_{\sun}$ PGC clouds of frozen  $10^{-6} M_{\sun}$ PFP
micro-brown-dwarf ``rogue planets''.  The present high resolution wide angle
HST/ACS observations of the Tadpole galaxy merger provide excellent evidence
of star wakes and YGCs triggered by objects merging through galactic
baryonic dark matter halos comprised of  PGC clumps of dark PFPs, as predicted
by HGT.  The observations contradict previous interpretations of Tadpole based
on cold-dark-matter halos, collisionless mergers, collisionless tidal tails,
and CDMHCC.

\section{Tadpole Observations}

Besides the eponymous tail, the most striking features of the very
high resolution
wide angle HST/ACS images of Tadpole are the two well defined luminous ovals
around VV29a which we interpret as star formation wakes in the dark
halo above and
below the disk of VV29a (Figure 2) rather than ordinary spiral arms in the
disk.  The outermost star-wake is labeled VV29c at the bottom of Fig. 2
because it is a direct continuation of the VV29b star-wake and is readily
extrapolated around VV29a to the embedded  galaxy VV29c.  The inner oval
pattern extends the center of the West Plume \citep{bri01} of stars that
apparently were triggered into formation by the merging galaxy components
VV29def, and appears to terminate at VV29e and some sort of AGN plasma jet
(labeled ``Polar Stream'' in the Briggs et al. 2001 inventory of structures
Fig. 13).

The inferred galaxy component VV29d leaves a dark dust wake plunging toward the
VV29a center in Figure 3.  VV29e also leaves a dark dust wake,
suggesting strong
turbulence and short lived massive stars were triggered by its passage.
Component VV29f entered the inner dark halo of VV29a closest to VV29c
and left no
stars small enough to remain luminous for more than the $ \approx 5
\times 10^{8}$
years since its passage.  The dust trail of VV29f leaves a narrow
shadow over the
bright oval of VV29de and terminates with the bright clump of star clusters
labeled VV29f in Fig. 2, maintaining a distance of about $1.5
\times 10^{20}$ from VV29c throughout their spiral descent toward VV29a.  This
interpretation is supported by the radio telescope observations \citep{bri01}
showing large integral HI (neutral hydrogen) column densities with velocity
opposite to that of the VV29a rotation and VV29c star wake.  The
VV29c and VV29f
objects are in front of and obscure the VV29b star-wake in the HST/ACS images,
contrary to the \citep{bri01} interpretation that these objects are $\approx 4
\times 10^{21}$ m in the background of VV29a using a frictionless tidal tail
scenario.

The edge of the VV29a baryonic dark matter halo is clearly shown by the high
resolution HST/ACS images, as in Figure 4 (dashed line at bottom right).
Above the boundary we see numerous young globular clusters,
especially along the
star-wakes of VV29cdef but also all around them, but no stars and no
YGCs below.
A gas patch with $0.6 \times 10^{24}$ H atoms m$^{-2}$ and with the
constant +100
km/s velocity measured for the gas of VV29b is shown in Fig. 1 of Briggs et al.
2001 at the beginning of the VV29cdef star-wakes shown in Fig. 4 for
the present HGT
interpretation.  Another patch with double the column density is shown to the
west of VV29a which can be identified with a patch of stars, presumably formed
by halo-halo tides, in the high resolution HST/ACS images.  Larger gas surface
densities up
to
$3.6
\times 10^{24}$ H atoms m$^{-2}$ were reported near the cluster of YGCs shown
near the top of Fig. 4, and even larger values at VV29a itself.  Such
high gas column
densities in star wakes only about $10^{20}$ m wide suggest H-He mass densities
$\approx 4 \times 10^{-23}$ kg m$^{-3}$ exist in the dark baryonic
halo of VV29a.  This
gives a total baryonic-dark-matter inner halo mass of order $10^{43}$ kg, which
is an order of magnitude  larger than
the mass of the central galaxy.  A similar conclusion results if one takes a
typical PGC separation in the inner halo (within $4 \times 10^{21}$ m) to be
about
$3
\times 10^{19}$ m based on the separation of bright objects taken to be YGCs
shown in the star-wake region of Fig. 4.

Keck telescope spectroscopy confirms our assumption that the
bright blue objects in the VV29b filamentary galaxy are YGCs.  Figure 5 shows
the slit location for the 42 YGC candidates identified \citep{tra03}.  The
brightest clump is described as a super-star-cluster (SSC), but with such a
large half light diameter ($\approx 10^{19}$ m) that it must either be unbound
or very ``cold''. The YGCs are aligned precisely (within
$\approx 1^o$) with the track of VV29c toward the beginning of its
spiral star-wake produced during capture by VV29a.  Our interpretation from
hydro-gravitational-theory is that the large clump of massive bright blue stars
(SSC) represents the first stars of to be triggered into formation
from PFPs in a clump of dark PGCs by the near passage of the dense central core
of VV29c.  Such massive stars likely form in PGC cores at large turbulent
Schwarz scales, reflecting maximal dissipation rates
$\varepsilon$ in gas evaporated from PFPs by VV29c as it passed
through a dense portion of the baryonic dark galaxy-halo of VV29a,
revealing a dark galaxy-halo dwarf-galaxy.  Frictional forces of the
observed gas explain why the apparent SSC is bound. Our star
formation process from pre-existing pre-stellar and
proto-cluster condensations (PGCs), formed much earlier in the universe
and constituting the BDM, also explains why such bright clusters of large
stars can form, without any evidence in the HST/ACS images of the giant 
($10^6 M_\sun$) molecular clouds (GMCs) ordinarily seen in Galactic regions
of star formation with ages of several million years (see the excellent
summary of embedded cluster formation in GMCs by Lada \& Lada, 2003), with
the 3 \--- 10 million year ages estimated by Tran et al. (2003).  From HGT,
the mysterious GMCs, identified as the site of most star
formation in our Galaxy \citep{LL2003}, are PGCs captured by the Galaxy
disk, with large stars of embedded clusters ($\approx 50 M_\sun$) reflecting
the large
$\varepsilon$ and
$L_{ST}$ values of star formation determined by strong resulting
turbulence.  New infrared observational capabilities such as SIRTF will
test our suggestion that star formation processes in dark PGCs and
GMCs are closely related.

For comparison with the large gas concentrations in VV29, Briggs et al. 2001
point out that no gas is detectable in the near companion elliptical galaxy MCG
09-26-54.  From HGT it is easy to understand why  gas
appears in merging galaxy systems like Tadpole as frozen PFP planetoids
evaporate, and not in quiescent galaxy cores like MCG 09-26-54 where
ambient gas freezes out on the same objects.  It is also easy to
understand the appearance of stars triggered by the passage of  galaxy
components through BDM halos if stars are formed by increased accretion rates
of PFPs caused by increased levels of gas friction.  It is impossible to
understand how collisionless, frictionless, tidal tail ejections of any
kind produced the high concentrations of gas observed in Tadpole, the spiral
star trails, the narrow filaments of YGCs without OGCs, or YGCs clustered in
straight lines pointing to the spiral star trails.

\section{Conclusions}

The wealth of morphological details in the high resolution HST/ACS  images
makes any frictionless tidal tail interpretation of the Tadpole system
untenable.  A tidal bridge to an invisible CDM halo
\citep{tre01} as the explanation of VV29b would require that the CDM halo
rotate precisely  with VV29a and that the VV29c star-wake near VV29a be
ignored.  Tidal forces in this CDM scenario should expel a symmetric trail of
stars on the opposite side of VV29a that is not observed.  Similarly, the
  suggestion \citep{bri01} that  VV29c entered from the Northwest rather than
the Southeast and is far in the  background moving away seems quite
implausible in view of the HST/ACS images that show VV29c and other companion
components VV29def are firmly embedded in,  and merging with, VV29a.  No
evidence of dust extinction caused by the spiral arms of VV29a is seen on the
face of VV29c in the HST/ACS images (eg Fig. 2) as one would expect if VV29c
were far in the background. CDM halos smaller than 100 kpc are ruled out by
hydro-gravitational-theory and cosmology, and are not supported by observations
\citep{sa02}.  Frictionless galaxy collisions and tidal tails are also ruled
out by HGT, and are not supported by the great quantities of gas observed in
Tadpole
\citep{bri01}.  Spectroscopic observations using the Keck telescope
\citep{tra03} show dense concentrations of young globular clusters aligned
closely in a star trail pointing toward the merging spiral star wake at the
central galaxy.  The 3-10 Myr ages of the YGCs found $2 \times 10^{21}$ m
from VV29a  proves they were formed in place and not 300 Myr later by unknown
processes after ejection at reasonable speed.  Their precise alignment with the
vector toward the merging spiral star-wake of VV29c around VV29a proves they
were formed as a wake and not ejected.

  From the HST/ACS images the Tadpole galaxy can best be interpreted as
a frictional galaxy merger producing baryonic-dark-matter star wakes,
young-globular-star-clusters, and dSphs (dwarf galaxies) from PFPs and PGCs
rather than any sort of collisionless galaxy merger flinging out tidal tails.
Gas that provides  the friction and triggers the star formation is produced by
evaporating frozen primordial-fog-particles in proto-globular-clusters
comprising the baryonic-dark-matter halos of the merging galaxies in Tadpole,
as in the BDM halos similarly revealed for the Mice and Antennae merging galaxy
systems. The location of the edge of the  VV29a baryonic dark matter halo is
clearly shown by the HST/ACS images giving a Tadpole BDM halo radius of $
\approx 4
\times 10^{21}$ m (120 kpc).  From the observed gas densities and velocities
\citep{bri01} and the YGC separation distances it appears that the
baryonic-dark-matter halo mass is at least an order of magnitude larger than
the mass of the luminous stars of  VV29a. Thus the HST/ACS images of Tadpole
support the existence of massive baryonic-dark-matter galaxy halos composed of
about a million dark proto-globular-clusters (PGCs) of dark
primordial-fog-particles (PFPs), as predicted by the hydro-gravitational theory
and  cosmology of Gibson 1996-2000 and confirmed by the quasar microlensing
interpretation of Schild 1996.

\acknowledgments

The authors are grateful for several excellent suggestions from the Editor and
the Referee.

\clearpage

\begin{figure}
         \epsscale{1.0}
         \plotone{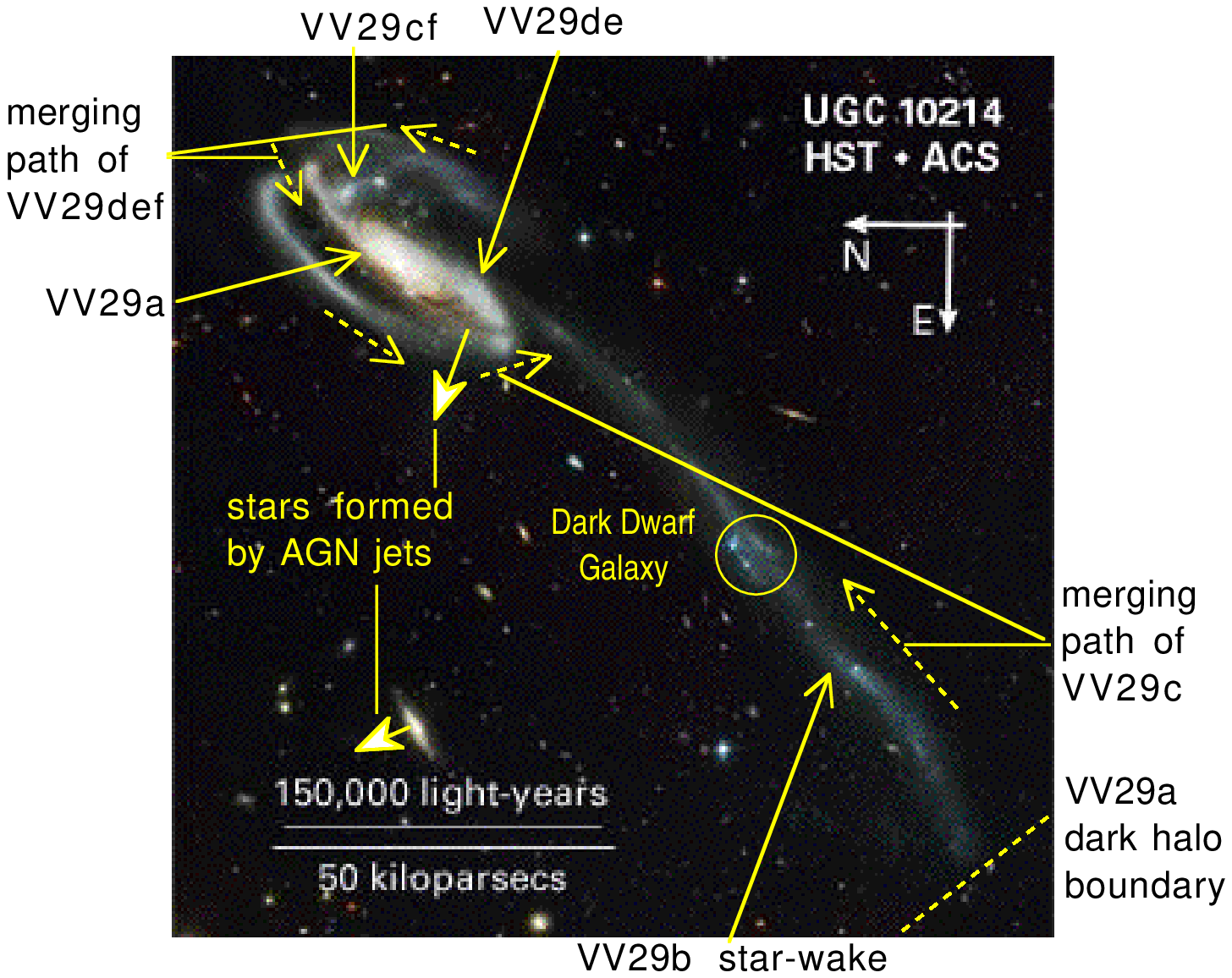}
         %%\plottwo{epsfile}{epsfile}
         \caption{Interpretation of HST/ACS Tadpole image (April 30, 2002
press release) using the Gibson 1996-2000 hydro-gravitational structure
formation  cosmology, where
the merging galaxy structures VV29cdef enter the VV29a dark halo at lower right
(dashed line) and merge along luminous trails of star formation triggered
by radiation and tidal forces acting on the PGCs and PFPs of the dark halo of
VV29a.  The VV29b galaxy is interpreted as a star-wake
of VV29cdef  merging galaxy
components in the VV29a dark baryonic halo, and not as any sort of
collisionless
tidal tail.  Star formation regions (visible in higher resolution images)
triggered in the baryonic-dark-matter by AGN jets of VV29e and a background
galaxy are shown by arrows.  A previously dark dwarf galaxy is revealed by a
straight row of closely spaced YGCs \citep{tra03} pointing toward the VV29c
entry star-wake (see Fig. 5). North is to the left and East is to the
bottom.}
         \end{figure}

\begin{figure}
         \epsscale{1.0}
         \plotone{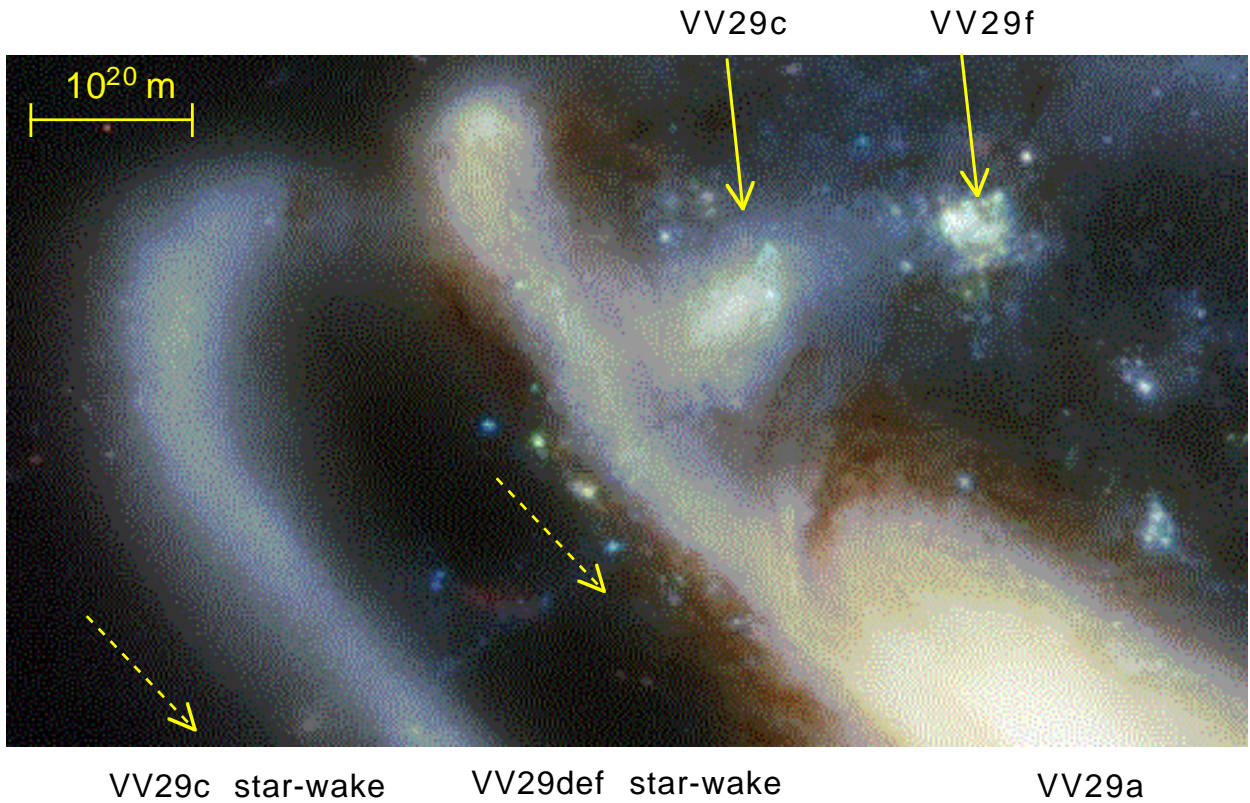}
         %%\plottwo{epsfile}{epsfile}
         \caption{Close-up of HST/ACS Tadpole image (April 30, 2002
press release) showing that the VV29c  blue dwarf galaxy and its companion
VV29f are merging with and embedded in the larger spiral galaxy VV29a.    Most
of the bright objects are probably young globular clusters formed from dark
PGCs by the merger event.  North is to the left and East is to the bottom.}
         \end{figure}

\begin{figure}
         \epsscale{1.0}
         \plotone{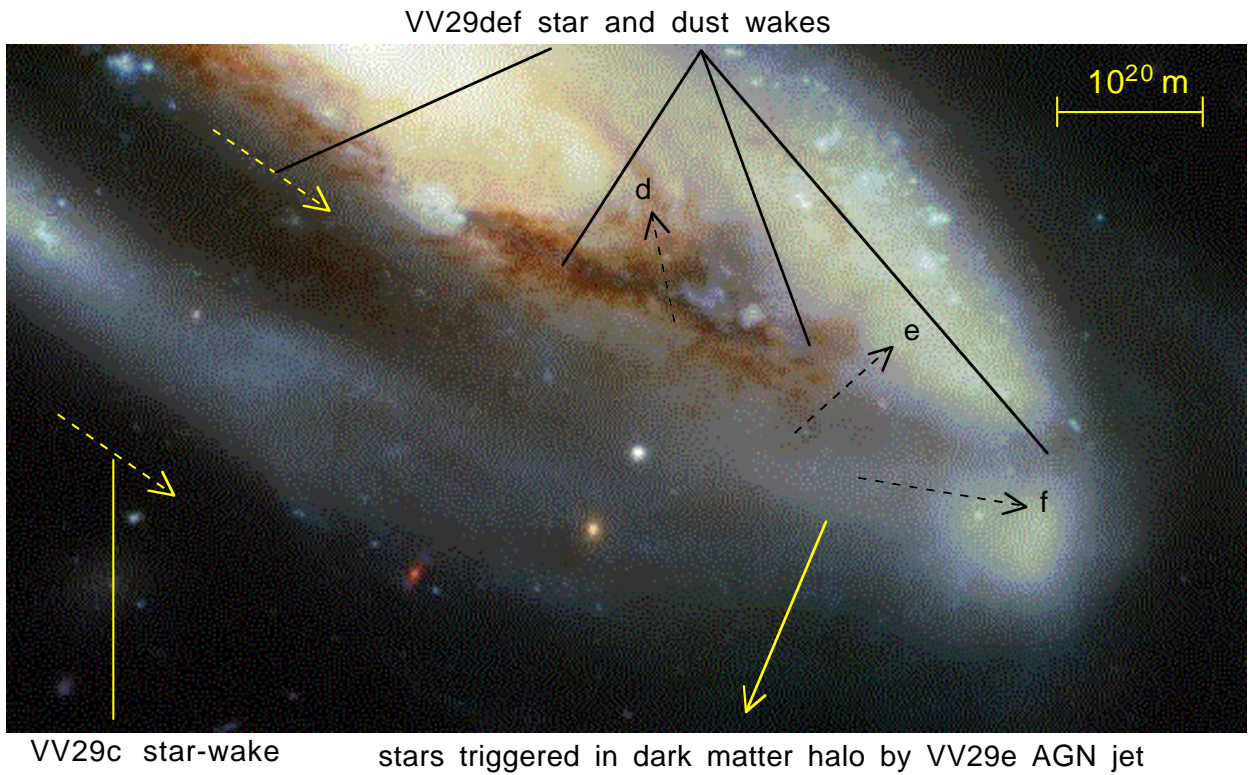}
         %%\plottwo{epsfile}{epsfile}
         \caption{Close-up of HST/ACS Tadpole image (April 30, 2002
press release) showing star wakes of merging dwarf galaxies VV29def.  VV29e
apparently contains an active galactic nucleus to explain the plume of stars
observed in the direction of the bottom arrow.  Note the dark VV29def
dust wakes
expected from massive stars formed in highly turbulent regions. North 
is to the left and East is
to the bottom.}
         \end{figure}

\begin{figure}
         \epsscale{.8}
         \plotone{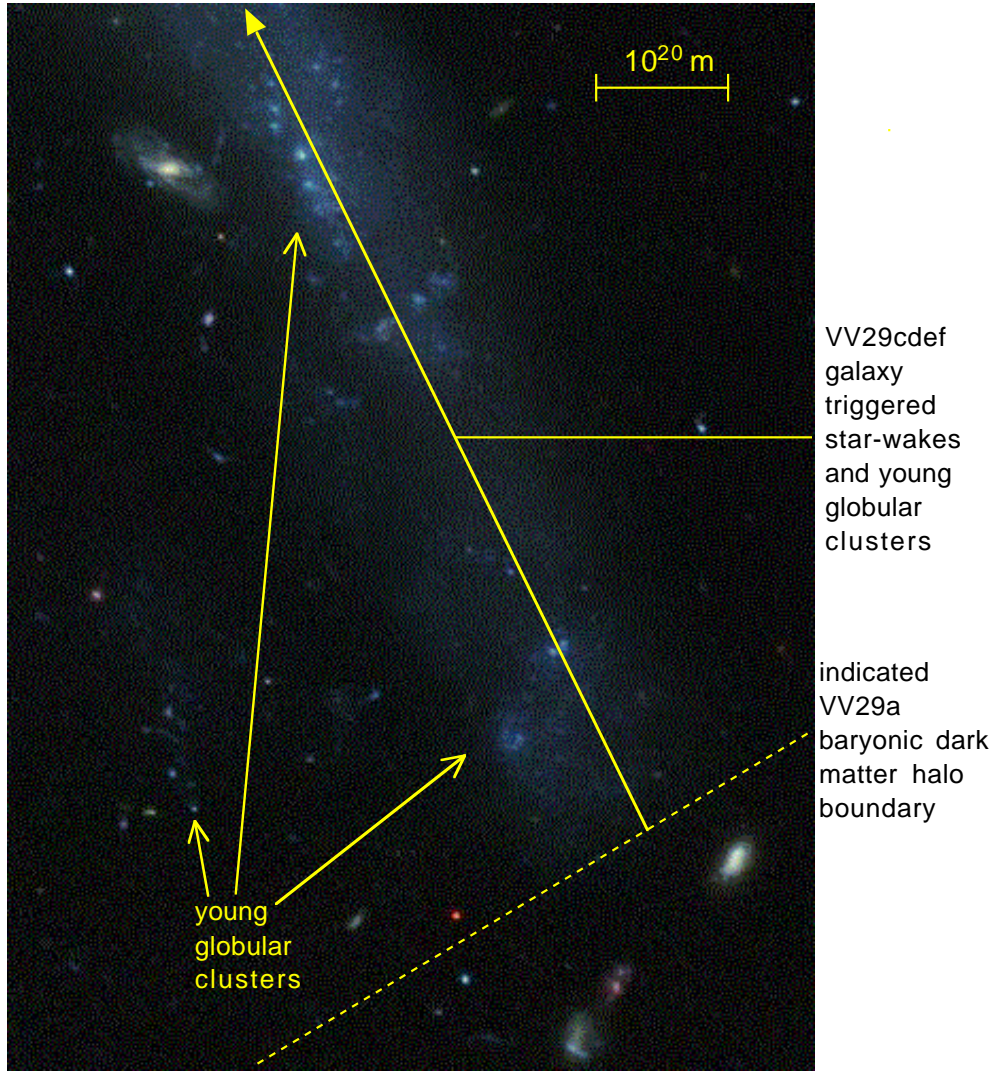}
         %%\plottwo{epsfile}{epsfile}
         \caption{Close-up of HST/ACS Tadpole image (April 30, 2002
press release) showing the edge of the baryonic dark matter halo of
galaxy VV29a
as indicated by the beginning of the star-wakes and young globular
clusters (YGCs)
triggered by the entry to the halo by galaxy components VV29cdef.  If the dark
halo is spherical with a homogeneous distribution of PGCs separated
by $\approx 3
\times 10^{19}$ m corresponding to the YGC separation shown, the dark halo mass
is about $10^{43}$ kg, with density $\rho \approx 4 \times 10^{-23}$ kg $\rm
m^{-3}$.  The dynamical mass of VV29a from its rotation velocity
within radius $7
\times 10^{20}$ m
\citep{bri01} is $\approx 10^{42}$ kg. North is to the left and East is
to the bottom.}
         \end{figure}

\begin{figure}
         \epsscale{1.0}
         \plotone{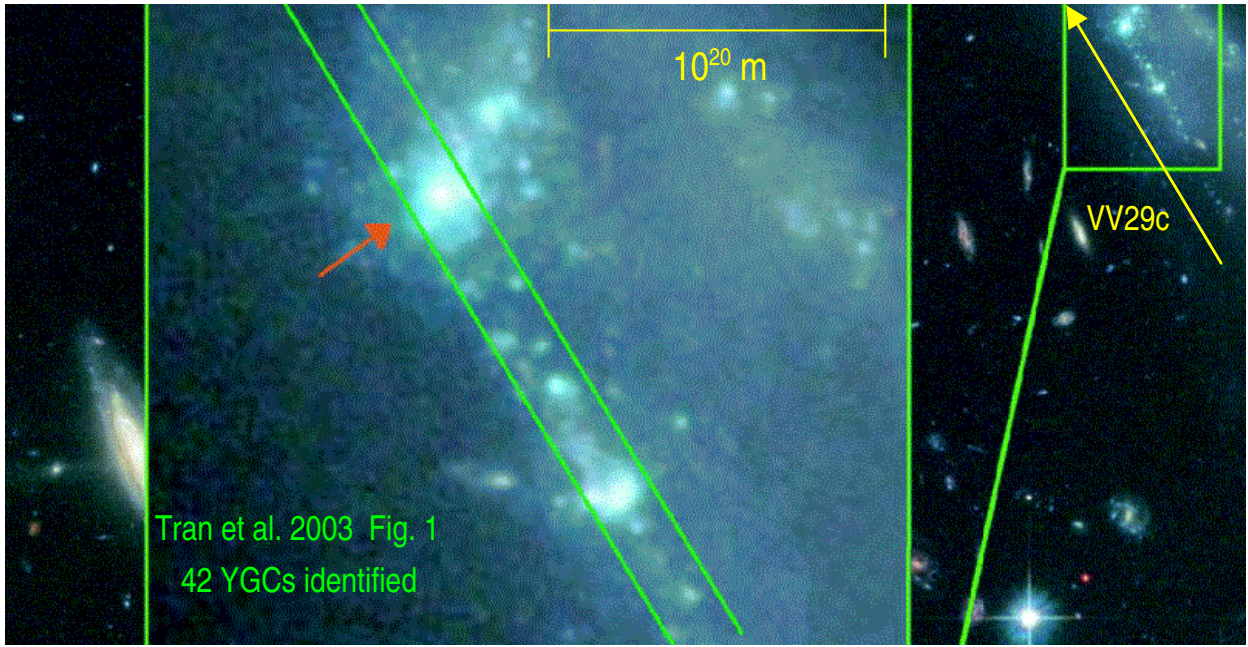}
         %%\plottwo{epsfile}{epsfile}
         \caption{Bright linear clump of YGCs in VV29b examined
spectroscopically by Tran et al. 2003 using the Keck telescope.  The 1 ''
Echellette slit used is shown with a loose super-star-cluster SSC
(arrow on left,  see text).  Ages of the 42 clusters identified range from 3-10
Myr.  The aligned clusters point precisely along the arrow (right, labeled) to
the beginning of the spiral star wake of VV29c in its capture, with
companions, by VV29a.  The large stellar mass density
$\rho
\approx 10^{-21}$ kg $\rm m^{-3}$ and the straightness  of the cluster row
suggests it is a star trail triggered by VV29c passing through an
$\approx 10^{39}$ kg dark dwarf galaxy
  in the VV29a BDM halo (see Fig. 1). North is to the left
and East is to the bottom.}
         \end{figure}

\clearpage

\begin{deluxetable}{lrrrrcrrrrr}
\tablewidth{0pt}
\tablecaption{Length scales of self-gravitational structure formation}
\tablehead{
\colhead{Length scale name}& \colhead{Symbol}           &
\colhead{Definition$^a$}      &
\colhead{Physical significance$^b$}           }
\startdata Jeans Acoustic & $L_J$ &$V_S /[\rho G]^{1/2}$& ideal gas pressure
equilibration
\\ Schwarz Diffusive & $L_{SD}$&$[D^2 /\rho G]^{1/4}$& $V_D$ balances $V_{G}$
\\  Schwarz Viscous & $L_{SV}$&$[\gamma \nu /\rho G]^{1/2}$& viscous force
balances gravitational force
  \\ Schwarz Turbulent & $L_{ST}$&$\varepsilon ^{1/2}/ [\rho G]^{3/4}$&
turbulence force  balances gravitational force
\\

Kolmogorov Viscous & $L_{K}$&$ [\nu ^3/ \varepsilon]^{1/4}$& turbulence
force  balances viscous force
\\

Ozmidov Buoyancy & $L_{R}$&$[\varepsilon/N^3]^{1/2}$& buoyancy force
balances turbulence force
\\

Particle Collision & $L_{C}$&$ m \sigma ^{-1} \rho ^{-1}$& distance between
particle collisions
\\

Hubble Horizon & $L_{H}$&$ ct$& maximum scale of causal connection
\\

%\cutinhead{This is a cut-in head}
%\sidehead{I am a side head:}

\enddata
\tablenotetext{a}{$V_S$ is sound speed, $\rho$ is density, $G$ is Newton's
constant, $D$ is the diffusivity, $V_D \equiv D/L$ is the diffusive velocity
at scale $L$, $V_G \equiv L[\rho G]^{1/2}$ is the gravitational velocity,
$\gamma$ is the strain rate,
$\nu$ is the kinematic viscosity,
$\varepsilon$ is the viscous dissipation rate, $N \equiv
[g\rho^{-1}\partial\rho/\partial z]^{1/2}$ is the stratification frequency,
$g$ is self-gravitational acceleration, $z$ is in the opposite direction
(up),
$m$ is the particle mass,
$\sigma$ is the collision cross section,  $c$ is light speed, $t$ is the age of
universe.}

\tablenotetext{b}{Magnetic and other forces (besides viscous and turbulence)
are negligible for the epoch of primordial self-gravitational structure
formation
\citep{gib96}.}

%% You can append references to a table using the \tablerefs command.

%\tablerefs{}
\end{deluxetable}

\clearpage

\begin{deluxetable}{lrrrrcrrrrr}
\tablewidth{0pt}
\tablecaption{Acronyms}
\tablehead{
\colhead{Acronym}& \colhead{Meaning}           &

\colhead{Physical significance}           }
\startdata

BDM & Baryonic Dark Matter&PGC clumps of PFPs from HGT
\\

CDM & Cold Dark Matter& an erroneous concept
\\

CDMHCC & CDM HCC&nested erroneous concepts
\\

HCC & Hierarchical Clustering Cosmology& an erroneous concept
\\

HCG & Hickson Compact Galaxy Cluster& Stephan's Quintet (SQ=HGC 92)
\\

HGT & Hydro-Gravitational Theory& corrects Jeans 1902
\\

ISM &Inter-Stellar Medium& mostly PFPs and gas from PFPs
\\

NBDM & Non-Baryonic Dark Matter&possibly neutrinos
  \\

OGC & Old Globular star Cluster& PGC forms stars at $t
\approx 10^{6}$ yr
\\

PFP&Primordial Fog Particle&planet-mass protogalaxy fragment
\\

PGC & Proto-Globular star Cluster& Jeans-mass protogalaxy fragment
\\

SSC & Super-Star Cluster&  a cluster of YGCs
\\

YGC & Young Globular star Cluster& PGC forms stars at $t
\approx$ now
\\

%\cutinhead{This is a cut-in head}
%\sidehead{I am a side head:}

\enddata

%% You can append references to a table using the \tablerefs command.

%\tablerefs{}
\end{deluxetable}

\end{document}